\documentstyle[prl,aps,epsf,epsfig,floats,twocolumn]{revtex}
\begin{document}
\draft
\twocolumn[\hsize\textwidth\columnwidth\hsize\csname
@twocolumnfalse\endcsname

\title{
\hfill{\small{DOE/ER/40762-218}}\\
\hfill{\small{UMD-PP-01-018}}\\[0.6cm]
Measuring the P-odd Pion-Nucleon Coupling $h_{\pi NN}^{(1)}$\\
in $\pi^+$-Photoproton Production Near the Threshold}
\author{Jiunn-Wei Chen and Xiangdong Ji}
\address{Department of Physics, University of Maryland,
College Park, MD 20742-4111}
\maketitle
\begin{abstract}
We show that $\overrightarrow{\gamma }p\rightarrow \pi ^{+}n$ in the
threshold region is an excellent candidate for measuring the leading
parity-violating pion-nucleon coupling $h_{\pi NN}^{(1)}$ to an uncertainty
of 20\% if it has a natural size from dimensional analysis. The conclusion is
based on a large unpolarized cross section, a new low-energy theorem for the
photon polarization asymmetry at the threshold $A_\gamma|_{{\rm th}} = \sqrt{%
2} f_\pi (\mu_p-\mu_n) h_{\pi NN}^{(1)}/g_A m_N \sim h_{\pi NN}^{(1)}/2$,
and its strong dominance at forward and backward angles in the threshold
region.
\end{abstract}
\medskip
{PACS numbers: 11.30.Er, 11.30.Rd, 13.60.Le.} 

]

\vspace{1cm}

Parity-violating, or P-odd, hadronic observables provide crucial information
about the physics of non-leptonic weak interactions in hadronic structures
and reactions. At low-energy, parity-violating hadronic interactions can be
systematically classified in the framework of effective field theories \cite
{KS,HBChPT,BKM}. At the leading order in chiral power counting, the most
important is the isovector P-odd pion-nucleon coupling $h_{\pi NN}^{(1)}$
which is responsible for the longest range part of the parity-violating $%
\Delta I=1$ $NN$ forces \cite{KS,DDH,AH}. In quantum chromodynamics (QCD), 
its value is dominated by
the $s$-quark contribution through neutral current interaction\cite{DSLS}. A
precise knowledge of $h_{\pi NN}^{(1)}$ not only is critical for
understanding the P-odd $NN$ force but will also shed important light on how
parity violation takes place in nonleptonic systems.

For many years, serious attempts have been made to measure $h_{\pi NN}^{(1)}$
from parity-violating processes (see \cite{AH,HH,Oers} for reviews). In
many-body systems, parity-violating effects can be enhanced by strong
correlations and have been detected experimentally. However, the theoretical
analyses have not yet been fully reliable. The disagreement in the
extraction of $h_{\pi NN}^{(1)}$ from $^{18}$F \cite{F18} and $^{133}$Cs 
\cite{Cs123} systems could be a reflection of poor understanding of
many-body physics. In few-body systems, the theory is under better control;
but the P-odd effects are generally small. While previous measurements could
not reach the required precision \cite{oldfew}, new experiments under way
are expected to improve significantly. These include $\overrightarrow{n}%
p\rightarrow d\gamma $ at LANSCE \cite{LANSCE}, $\gamma d\rightarrow np$ at
Jefferson Lab (JLab)\cite{JLab}, and the rotation of polarized neutrons in helium at NIST 
\cite{oldfew}. Finally, in the single nucleon systems, new P-odd
observables in Compton scattering on the proton were recently proposed to
determine $h_{\pi NN}^{(1)}$ \cite{BS}. The process is
theoretically ``clean'', however the experimental feasibility is marginal because of
the small total cross section and P-odd asymmetries.

In this paper, we show that the polarized photon asymmetry in $%
\overrightarrow{\gamma }p \rightarrow n\pi ^{+}$ at the threshold region is
an execellet candidate to measure $h_{\pi NN}^{(1)}$. We derive a low-energy
theorem for the asymmetry at the pion-production threshold in the chiral 
limit: $A_\gamma|_{{\rm %
th}} = \sqrt{2}f_\pi(\mu_p-\mu_n) h_{\pi NN}^{(1)}/g_A m_N\sim h_{\pi
NN}^{(1)}/2$. A leading-order (LO) calculation in heavy-baryon chiral
perturbation theory (HB$\chi$PT) shows that the result is modified only mildly
by higher partial waves, particularly at forward and 
backward angles, and chiral corrections from the finite pion mass 
and momentum in the threshold region up to photon energy 
$E_\gamma\sim 200$ MeV. With a total
cross section $\sim 100 \mu$b and the expected asymmetry $\sim 2\times
10^{-7}$, the experiment is feasible at existing laboratories such as 
JLab.
Theoretical studies of the same process have been carried out before by
Woloshyn \cite{Woloshyn} and by Li, Henley and Hwang \cite{LHH} in the
framework of meson exchange models. In particular, Ref. \cite{LHH} has
already noted the dominance of the $h_{\pi NN}^{(1)}$-type P-odd coupling
in the asymmetry near the threshold. The present analysis sharpens the
finding by deriving the low-energy theorem and defending its dominance in
the threshold region using the modern theoretical tool---HB$\chi$PT
\cite{HBChPT,BKM}.

We are interested in the following two-body process, 
\begin{equation}
\overrightarrow{\gamma }\left( q^{\mu };\epsilon ^{\mu }\right)
+p(P_{i}^{\mu })\rightarrow \pi ^{+}(k^{\mu })+n(P_{f}^{\mu })\ ,
\end{equation}
where $q^{\mu }=\left( \omega ,{\bf q}\right) $, $P_{i}^{\mu }$, $k^{\mu
}=\left( \omega _{\pi },{\bf k}\right) $, and $P_{f}^{\mu }$ are the
center-of-mass four-momenta of photon, proton, pion and neutron,
respectively, and $\epsilon ^{\mu }$ is the photon polarization vector. In
the threshold region, the pion and photon as well as the nucleon momenta
are much smaller than the chiral symmetry breaking scale 
$\Lambda_\chi \sim 4 \pi f_{\pi }\sim 1$ GeV; therefore chiral perturbation theory 
($\chi $PT) is a useful tool in making theoretical analyses \cite{BKM}. 
When the nucleon is explicitly involved, a natural scheme for 
systematic power counting is to treat its
mass as a heavy scale as $\Lambda_\chi$, and thus HB$\chi$PT 
\cite{HBChPT}. In addition, since the delta-resonance is 
only $300$ MeV heavier than the nucleon
(order $1/N_{c}$ in QCD with a large number of $N_{c}$ colors)
and is strongly coupled to the latter through electromagnetic exitations,
it is sensible to extend HB$\chi $PT to include the resonance as dynamical
degrees of freedom and to treat the mass difference $\Delta=m_\Delta -m_N$ 
as a small parameter \cite{Hemmert}. The $SU(2)_{L}\times U(1)$ symmetry 
structure of electroweak
interactions can be incorporated with the weak boson exchange described by
contact interactions and the photon kept as dynamical degrees of freedom.

The unpolarized $\gamma p\rightarrow \pi ^{+}n$ reaction at the threshold
represents a classical example of the successes of effective theory
ideas. Simply relying on the symmetry properties of the strong interactions,
Kroll and Ruderman made a prediction in 1954 on the $s$-wave scattering
length in the chiral limit \cite{KR}. Away from this limit, the corrections
have been successfully studied using effective field theories. A first
analysis of the reaction in $\chi $PT was made by Bernard et al. 
\cite{BKMII}, who found that the one-loop correction to the 
tree-order threshold $s$-wave amplitude ($E_{0+}$) is 
insignificant. A more detailed study of partial
waves in the framework of HB$\chi $PT has recently been made 
by Fearing et al. \cite{Fearing}
, who found that the $p$-wave multiples at the threshold are well described
by the leading (${\cal O}(p)$) plus next-to-leading $(%
{\cal O}(p^{2}))$ order calculations. For example, $M_{1+}$, $M_{1-}$, 
and $E_{1+}$ multiples are $-4.7$, $9.4$, and $4.7$ in unit 
$10^{-3}/m_{\pi }^{3}{}^{+}$ at ${\cal O}(p)$. At order ${\cal O}(p^{2})$, 
the results are $-7.7$, $5.6$,
and $5.1$ which compare favorably with $-9.6$, $6.1$%
, and $4.9$ from a dispersion-theory analysis of experimental data 
\cite{HDT}.

For the process to be useful in studying nonleptonic parity-violating
interactions, the cross section must be large enough to yield 
a sufficient number of events. Because of the severe phase 
space suppression at 
the threshold, we need to establish an {\it extended} threshold 
region in which the effective theory description remains effective 
and, at the same time, the cross section is appreciable. For this purpose, 
we consider the result of HB$\chi $PT at leading order. 
The parity-conserving T-matrix depends on the four amplitudes, 
\begin{eqnarray}
T^{PC} &=&N^{\dagger }\left[ i{\cal A}_{1}\,{\bf \sigma }\cdot {\bf \epsilon 
}+i{\cal A}_{2}{\bf \sigma }\cdot \widehat{{\bf q}}{\bf \ }\,{\bf \epsilon }%
\cdot \widehat{{\bf k}}+i{\cal A}_{3}{\bf \sigma }\cdot \widehat{{\bf
k}}{\bf \ }\,%
{\bf \epsilon }\cdot \widehat{{\bf k}}\right.   \nonumber \\
&&\left. +{\cal A}_{4}{\bf \epsilon }\cdot \,\widehat{{\bf q}}\times 
\widehat{{\bf k}}\right] N\ ,  \label{tpc}
\end{eqnarray}
where $N$ is the proton Pauli spinor, ${\bf \sigma }$ is the Pauli
matrix vector, $\widehat{{\bf q}}$ and $\widehat{{\bf k}}$ 
are the unit vectors
in the ${\bf q}$ and ${\bf k}$ directions. At leading order in HB$\chi $PT,
${\cal A}_{1}=eg_{A}/\sqrt{2}f_{\pi }$, ${\cal A}_{2}={\cal A}_{1}\omega |%
{\bf k}|/q\cdot k$, ${\cal A}_{3}=-{\cal A}_{1}{\bf k}^{2}/q\cdot k$, and ${\cal A}%
_{4}=0$ \cite{Fearing}. The resulting differential cross section is 
\begin{eqnarray}
\frac{d\sigma }{d\Omega } &=&\frac{\alpha _{{\rm em}}g_{A}^{2}}{8\pi f_{\pi
}^{2}}\frac{m_{N}^{2}}{S}\frac{\left| {\bf k}\right| }{\omega }{\cal G}\
,~~~~  \nonumber \\
{\cal G} &=&1-\frac{\sin ^{2}\theta \left| {\bf k}\right| ^{2}}{q{\bf \cdot }%
k}\left[ 1-\frac{\left( {\bf q-k}\right) ^{2}}{2q{\bf \cdot }k}\right] \ ,
\label{g}
\end{eqnarray}
where $\theta $ is the angle between $\,\widehat{{\bf q}}$ and $\widehat{%
{\bf k}}$, and $S=(q+P_{i})^{2}$. A comparison between data 
\cite{totdata} and the
integrated cross section is shown in the upper graph of Fig.~1 as 
a function of the photon
energy $E_{\gamma }$ in the lab frame. The leading-order result 
describes the data (which have a considerable variation themselves)
within 10\% up to $E_{\gamma }=200$ MeV. The difference 
indicates the size of the higher-order corrections expected 
of HB$\chi $PT
and the level of convergence of the chiral expansion. 
According to the figure, we {\it define} the threshold region in terms of the laboratory 
photon energy from the threshold to 200 MeV. In the lower graph, 
we show the angular distributions of the pions in the center-of-mass
frame and the data which show the largest deviation from the theory 
by about 20\% at 200 MeV and backward angles. 


\begin{figure}[t]
\begin{center}
\epsfxsize=7.25cm
\centerline{\epsffile{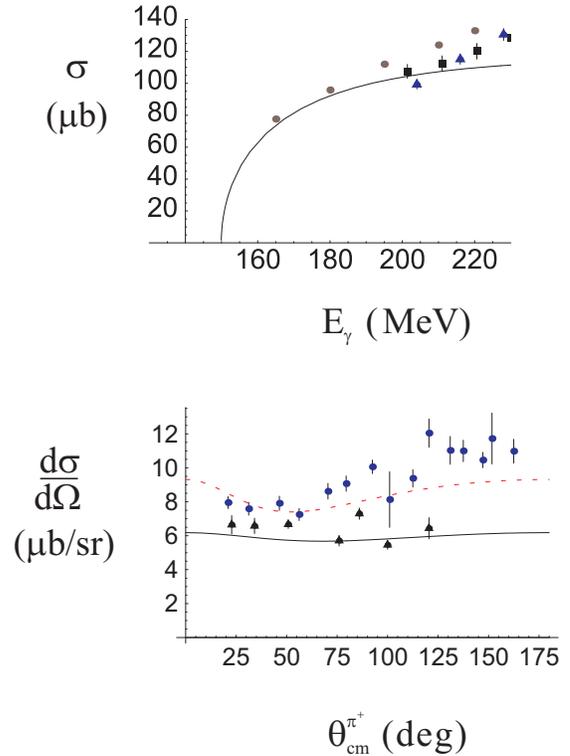}}
\end{center}
\caption{ Upper graph: $\protect\gamma p\rightarrow \protect\pi ^{+}n$ cross
section shown as a function of the photon energy in the laboratory frame.
The solid curve is the leading-order HB$\protect\chi$PT prediction
and the data shown are taken from Ref.\protect\cite{totdata}. Lower graph: 
the $\protect\pi^{+}$
angular distribution in the center-of-mass frame. The solid and 
short-dashed curves, and the corresponding data \protect\cite{HDT,totdata}, 
triangles and solid circles, are for $E_{\protect\gamma}=$ 165 
and 200 MeV, respectively.}
\label{fig:PCsig}
\end{figure}

Now we turn to parity-violating effects in the process. To calculate P-odd
observables, we need to extend chiral perturbation theory to 
include non-leptonic weak interactions. A systematic construction of 
the P-odd effective chiral lagrangian has been undertaken 
in Ref. \cite{KS}. To ${\cal O}(p^0)$ (we
choose to ignore the weak coupling in power counting), it has one term, 
\begin{equation}
{\cal L}^{PV}=-ih_{\pi NN}^{(1)}\pi ^{+}p^{\dagger }n+h.c.+\cdots \ , 
\label{Lw}
\end{equation}
where the ellipses denote terms with more pion fields and derivatives, and
the phase convention is taken from Refs.\cite{vanKolck}. By matching
onto four-quark interactions, $h_{\pi NN}^{(1)}$ was found to be dominated
by $s$-quark contributions, $\left| h_{\pi NN}^{(1)}\right| \sim G_{F}F_{\pi
}\Lambda _{\chi }/\sqrt{2}\sim 5\times 10^{-7}$ \cite{KS}. 
This estimation is
consistent with the ``best value'' obtained in Ref.\cite{DDH} and close to
a result \cite{HHK} from QCD sum rules. On the other hand, a 
recent calculation in the SU(3)
Skyrme model yields $h_{\pi NN}^{(1)}\sim 0.8$-$1.3$ $\times 10^{-7}$\cite
{MW}.

\begin{figure}[t]
\begin{center}
\epsfxsize=7.25cm
\centerline{\epsffile{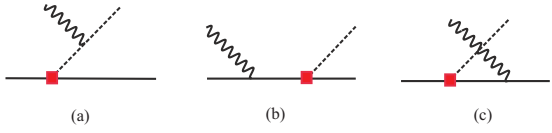}}
\end{center}
\caption{
Feynman diagrams contributing to the parity-violating amplitudes at
LO (${\cal O}(1)$) and NLO (${\cal O}(p)$) in 
$\protect\overrightarrow{\protect\gamma} p 
\protect\rightarrow \protect\pi^{+}n$. }
\label{PVLO}
\end{figure}

To the next-to-leading order (NLO) (${\cal O(}p{\cal )}$) in chiral 
expansion, the relevant Feynman diagrams for the P-odd $\gamma
p\rightarrow \pi ^{+}n$ process are shown in Fig. 2. The resulting
T-matrix can be expressed in terms of two amplitudes,
\begin{equation}
T^{PV}=N^{\dagger }\left[ i{\cal F}_{1}\,\widehat{{\bf k}}\cdot {\bf %
\epsilon }+{\cal F}_{2}{\bf \sigma }\cdot {\bf \epsilon }\times \,\widehat{%
{\bf q}}\right] \ ,
\end{equation}
where
\begin{equation}
{\cal F}_{1}=-\frac{eh_{\pi NN}^{(1)}\left| {\bf k}\right| }{q{\bf \cdot }k}%
\ ,\ {\cal F}_{2}=\frac{eh_{\pi NN}^{(1)}}{2m_{N}}\left[ \mu _{p}-\left( 
\frac{\omega }{\omega _{\pi }}\right) \mu _{n}\right] \ .
\end{equation}
P-odd observables can now be constructed from the interference between $%
T^{PV}$ and $T^{PC}$. The leading single-spin asymmetry arises from the
interference between ${\cal A}_{1-3}$ and ${\cal F}_{1}$, and is dependent 
on the proton polarization. Because of technical difficulties 
with a large volume, high-density polarized hydrogen target, 
an experimental 
measurement of this asymmetry is not within sight. Therefore, 
in the following we focus on the photon
helicity-flip asymmetry which comes in at NLO from the interferences 
between ${\cal A}_{1-3}$ and ${\cal F}_{2}$ and between 
${\cal A}_{4}$ and ${\cal F}
_{1}$. ${\cal A}_{4}$ in HB$\chi $PT is found nonvanishing at NLO 
and is 
\begin{eqnarray}
{\cal A}_{4} &=&\frac{eg_{A}\left| {\bf k}\right| }{2\sqrt{2}f_{\pi }m_{N}}
\left[ \mu _{p}-\left( \frac{\omega }{\omega _{\pi }}\right) \mu _{n}\right] 
\nonumber \\
&&-\frac{2eg_{\pi N\Delta }G_{1}\left| {\bf k}\right| }{9\sqrt{2}f_{\pi
}m_{N}}\left( \frac{\omega }{\omega -\Delta }+\frac{\omega }{\omega _{\pi
}+\Delta }\right) \ ,
\end{eqnarray}
where the delta-resonance contribution has been included explicitly. $G_1$
is the M1 transition moment between the nucleon and delta, and 
$g_{\pi N\Delta}$ is the $\pi$-$N$-$\Delta$ coupling.

\begin{figure}[t]
\begin{center}
\epsfxsize=7.25cm
\centerline{\epsffile{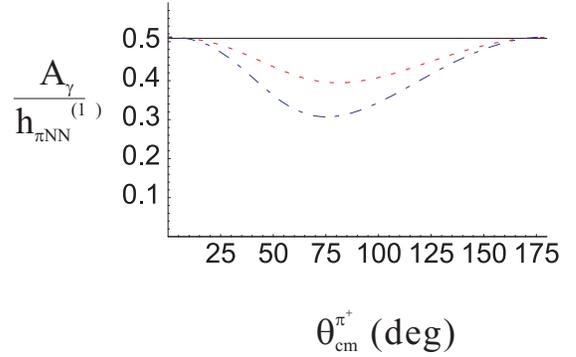}}
\end{center}
\caption{The photon-helicity asymmetry $A_{\protect\gamma }$
in unit $h_{\protect\pi NN}^{(1)}$. The solid line is the low-energy
theorem in Eq. (9), and the short-dashed and dash-dotted 
lines are for $E_{\protect}\protect\gamma =$180 and 200 MeV,
respectively.}
\label{fig:asym}
\end{figure}

More explicitly, the photon helicity
asymmetry $A_{\gamma }( \omega ,\theta )$$=$$(d\sigma (
\lambda _{\gamma }=+1) -d\sigma ( \lambda _{\gamma }=-1))$$/$$(
d\sigma ( \lambda _{\gamma }=+1) +d\sigma ( \lambda _{\gamma
}=-1) )$ at the leading order in HB$\chi$PT is   
\begin{eqnarray}
A_{\gamma }(\omega ,\theta ) &=&\frac{\sqrt{2}h_{\pi NN}^{(1)}f_{\pi }}{%
g_{A}m_{N}{\cal G}}\left\{ \left[ \mu _{p}-\left( \frac{\omega }{\omega
_{\pi }}\right) \mu _{n}\right] \left( 1-\frac{\sin ^{2}\theta {\bf k}^{2}}{q%
{\bf \cdot }k}\right) \right.   \nonumber \\
&&\left. +\frac{2}{9}\frac{g_{\pi N\Delta }G_{1}\sin ^{2}\theta {\bf k}^{2}}{%
g_{A}q{\bf \cdot }k}\left( \frac{\omega }{\omega -\Delta }+\frac{\omega }{%
\omega _{\pi }+\Delta }\right) \right\} 
\end{eqnarray}
where ${\cal G}$ is given in Eq. (\ref{g}). Although the result formally 
depends on the NLO amplitude ${\cal A}_4$, it is dominated in the
threshold
region by the ``beat'' between the parity-violating amplitude ${\cal F}_2$
and the leading-order parity-conserving amplitudes ${\cal A}_{1,2,3}$ 
which have already been tested in Fig. 1. 
Right at the threshold $|{\bf k}|=0$, only the $s$-wave $\pi ^{+}n$ 
final-state contributes; we find the equivalent of the 
Kroll-Ruderman theorem for the P-odd photon-helicity asymmetry, 
\begin{equation}
A_{\gamma }\left( \omega _{{\rm th}},\theta \right) =\frac{\sqrt{2}f_{\pi
}(\mu _{p}-\mu _{n})}{g_{A}m_{N}}h_{\pi NN}^{(1)}\ ,
\end{equation}
which depends only on the chiral symmetry. Plugging in the known physical
quantities, the coefficient of $h_{\pi NN}^{(1)}$ is 0.52. So the asymmetry
has the same size as $h_{\pi NN}^{(1)}$ and of order $10^{-7}$. 
In Fig. \ref{fig:asym}, we show the
angular dependence of the leading-order $A_{\gamma }$ at $E_{\gamma }=$180,
200 MeV (corresponding to the center-of-mass energy 
$\omega =138,168$ MeV), together with the low-energy theorem. 
At the forward and backward angles, we see hardly any deviation 
from the threshold result. Only near $\theta =90^{\circ }$
at $E_{\gamma }=200 $ MeV does the modification from high partial 
waves become significant (less than 40\%). 

Will the above result be changed significantly when going to higher orders
in HB$\chi$PT? A complete answer to the question
requires a systematic study of the contribution at the next order which we
will communicate in a separate publication \cite{chenji2}. Here we just
present a few qualitative arguments why it is unlikely that
the higher-order corrections ruin the leading-order 
relation between $A_{\gamma }$ and $h_{\pi
NN}^{(1)}$ in the threshold region. Because the parity-conserving 
amplitudes are dominated by the leading order, we know at least one
class of corrections---the interference between the next-to-next-to-leading order (NNLO) $T^{PC}$ with LO $%
T^{PV}$---is small. The second class of
corrections is an interference between NLO $T^{PC}$and NLO $T^{PV}$. 
No loop calculations are involved here and all couplings 
except $h_{\pi NN}^{(1)}$ are known. The size of the correction
will follow the canonical power counting, i.e.  of order $%
{\cal O}(\epsilon /m_{N})$, where $\epsilon$ stands for $m_\pi$, 
$\omega$, $\omega_\pi$, and $\Delta$. The last class involves an interference between
LO $T^{PC}$and NNLO $T^{PV}$ amplitudes; the latter contains 
one-loop integrals as well as tree contributions from new P-odd 
effective couplings. The following is an example of P-odd 
interactions at NNLO, 
\begin{equation}
{\cal L}^{PV}=\frac{eh_{\gamma \pi NN }}{m_{N}^{2}}\overline{p}\left[ S^{\mu
},S^{\nu }\right] \pi ^{+}nF_{\mu \nu }+i\frac{e\widetilde{G}}{m_{N}}%
\overline{\Delta ^{+}}^{\mu }v^{\nu }F_{\nu \mu }p.
\end{equation}
While the one-loop integrals are not expected to yield large corrections,
the magnitude of the new couplings is unknown. Since an unnatural
size of couplings in effective theory usually arises 
from new physics, we do not expect this to happen here from 
our experience with the corresponding parity-conserving amplitudes. 
This of course can be tested by the $\theta $ 
dependence of the asymmetry. In short, we expect the higher-order
corrections to Eq. (8) is ${\cal O}(\epsilon/m_N)$, namely, about
20\%.
  
Finally, we briefly comment on the experimental feasibility for measuring the
polarization asymmetry in $\overrightarrow{\gamma }p\rightarrow \pi ^{+}n$.
To overcome statistics, a large number of events ($\sim 10^{14}$) are
needed. This requires a luminosity of order 10$^{37}$/(cm$^{2}\sec $)
which is reasonable with the current technology and facilities such as
JLab. With a total cross section 
$\sim $100 $\mu b$=10$^{-28} $cm$%
^{2}$, the $\pi ^{+}$ production rate is 10$^{8} $/sec$\cdot $rad. 
Thus $\sim $10$^{6}$ sec of beam time will yield the required number of events.
The challenge, however, could be 10$^{8}\pi ^{+}$/sec detection.

In conclusion, we have shown that parity-violating $\overrightarrow{\gamma }%
p\rightarrow \pi ^{+}n$ is a theoretically clean and experimentally feasible
process to measure $h_{\pi NN}^{(1)}$. Near the threshold region, the size
of the photon helicity asymmetry is estimated to be $\sim 2\times 10^{-7}$
for an expected magnitude of $h_{\pi NN}^{(1)}$. 
Assuming a luminosity of 10$^{37}$/(cm$^{2}\sec $), 
$h_{\pi NN}^{(1)}$ can be measured to an accuracy of $10^{-7}$ 
in a few months of running. Similar results for pion 
electroproduction will be published separetely \cite{chenji2}.

Note added in proof: After this paper was submitted for publication,
Zhu {\it et al.} published a preprint presenting a calculation of
the NLO corrections \cite{zhu}. We have submitted a comment about 
their paper to e-archive \cite{zhu}.
\acknowledgements

We thank D. Beck, E. Beise, E. Henley, R. McKeown, R. Suleiman, and
B. Wojtsekhowski 
for discussions on the experimental issues.
This work is supported in part by the U.S. Dept. of Energy
under grant No. DE-FG02-93ER-40762.

\end{document}